\documentclass[showpacs,preprintnumbers,amsmath,amssymb,superscriptaddress,preprint]{revtex4}

\usepackage{graphicx}
\usepackage{dcolumn}
\usepackage{bm}

\newcommand{\figwidth}{0.5\textwidth}
\newcommand{\im}{\mathbf{Im}}
\newcommand{\re}{\mathbf{Re}}

\begin{document}

\title{High frequency thermoelectric response in correlated electronic systems}

\author{Wenhu Xu}
\affiliation{
Department of Physics and Astronomy, Rutgers University \\
136 Frelinghuysen Rd., Piscataway, NJ 08854, USA}
\author{C\'edric Weber}
\affiliation{Cavendish Laboratories, Cambridge University, JJ Thomson Avenue, Cambridge, UK}
\author{Gabriel Kotliar}
\affiliation{
Department of Physics and Astronomy, Rutgers University \\
136 Frelinghuysen Rd., Piscataway, NJ 08854, USA}

\date{\today}

\begin{abstract}
We derive a general formalism for evaluating the high-frequency
limit of the thermoelectric power of strongly correlated materials,
which can be straightforwardly implemented in available first principles LDA+DMFT programs.
We explore this formalism using model Hamiltonians and we investigate the validity of
approximating the static thermoelectric power $S_0$, by its
high-temperature limit, $S^*$. We point out that the behaviors of $S^*$ and
$S_0$ are qualitatively different for a correlated Fermi liquid near the Mott transition,
when the temperature is in the coherent regime. When the temperature is well above the coherent
regime, e.g., when the transport is dominated by incoherent excitations, $S^*$ provides a good estimation of $S_0$.
\end{abstract}

\pacs{71.10.-w, 71.15.-m, 72.15.Jf}
\maketitle

\section{Introduction}\label{sec:intro}

Thermoelectric energy harvesting, i.e. the transformation of
waste heat into usable electricity, is of great current interest.
The main obstacle is the low efficiency of materials for
converting heat to electricity \cite{mahan_ssp, mahan_phystoday}.
Over the past decade, there has been a renewed interest
on thermoelectric materials, mainly driven by experimental results \cite{snyder}.

Computing the thermoelectric power (TEP) in correlated systems is a
highly non-trivial task and several approximation schemes have
been used to this intent. The well-known Mott-Heikes
formula\cite{PhysRevB.13.647} gives an estimate of the high temperature
limit of TEP \cite{PhysRevB.62.6869} in the strongly correlated regime.
A generalized Boltzmann approach including vertex corrections has been
developed in Ref.~\cite{PhysRevB.67.014408} and applied to several materials.
Thermoelectric transport at intermediate temperature was carefully investigated
in the context of single-band and degenerate Hubbard Hamiltonians,
by dynamical mean field theory (DMFT) \cite{PhysRevLett.80.4775, PhysRevB.65.075102}.
Kelvin formula was also revisited for various correlated models
in Ref.~\cite{PhysRevB.82.195105} very recently.

The high frequency (AC) limit provides another interesting insights to
gain further understanding of the thermoelectric transport in correlated materials,
and is the main interest of this work. The thermopower in the high frequency
limit of a degenerate Hubbard model near half-filling was considered in Ref.~\cite{PhysRevB.65.075102},
where the authors generalize the thermoelectric response to finite frequencies in the high temperature
limit. The same limit was studied recently by Shastry and collaborators, who have developed a formalism
for evaluating the AC limit of thermoelectric response using high
temperature series expansion and exact diagonalization. The methodology was
applied to a single band t-J model on a triangular lattice\cite{0034-4885-72-1-016501,PhysRevB.73.085117}.
The authors pointed out that the AC limit of TEP ($S^*$) is simple enough that
it can be obtained by theoretical calculations with significantly less
effort, while still provides nontrivial informations of the thermoelectric properties, and give an estimation of
the trend of $S_0$.

In this work, we investigate the high frequency limit of TEP, $S^*$,
by deriving an exact formalism in the context of a general multi-band model with local interactions.
We show that $S^*$ is determined by the bare band structure and the
single-particle spectral functions. The relation between the conventional TEP, i.e., obtained
at zero frequency ($S_0$) and the AC limit $S^*$ is
discussed from general arguments on the single particle properties of correlated systems at low and
high temperatures. The analytical derivation of $S^*$ is compared
with the frequency dependent thermopower of the one band Hubbard model, solved by
dynamical mean field theory (DMFT) on the square and triangular lattices.
The formalism derived in this work can be conveniently implemented into first-principles
calculations of realistic materials, such as in the LDA+DMFT framework \cite{RevModPhys.78.865, held}.

This paper is organized as follows. In Sec.~\ref{subsec:gen_form},
general formalism of dynamical thermoelectric transport coefficients
is summarized to define the notation. In Sec.~\ref{subsec:sum_rules},
exact formulae to evaluate $S^*$ are derived for a general tight-binding
model with local interactions. In Sec.~\ref{sec:dmft}, we apply the
formalism to one-band Hubbard model on square and triangular lattice.
The low and high temperature limit behaviors of $S^*$ are discussed and
compared to those of $S_0$. Numerical results are presented in Sec.~\ref{sec:numerical}.
Sec.~\ref{sec:conclusion} summarizes the paper.

\section{Dynamical thermoelectric transport functions and high-frequency limit of thermopower}\label{sec:formal}

\subsection{General formalism}\label{subsec:gen_form}

Electrical current can be induced by gradient of electrical
potential and temperature. The phenomenological equations for
static(DC limit) external fields are\cite{mahan_many}
\begin{eqnarray}
J^x_1=L^{xx}_{11}\left(-\frac{1}{T}\nabla_x
\tilde{\mu}\right)+L^{xx}_{12}\left(\nabla_x\frac{1}{T}\right), \label{eq:linear_J1}\\
J^x_2=L^{xx}_{21}\left(-\frac{1}{T}\nabla_x
\tilde{\mu}\right)+L^{xx}_{22}\left(\nabla_x\frac{1}{T}\right). \label{eq:linear_J2}
\end{eqnarray}
We only consider the longitudinal case. $J_1^x$ and $J_2^x$ are $x-$
component of particle and heat current, respectively. $\nabla_x
\tilde{\mu}$ and $\nabla_x\frac{1}{T}$ are generalized forces driving
$J^x_1$ and $J^x_2$. $\tilde{\mu}=\mu-eV$, in which $\mu$
is chemical potential and $V$ is the electric potential.
$L^{xx}_{ij}$ are transport coefficients. We follow the definition in Ref.~\cite{mahan_many},
which explicitly respects the Onsager relation, $L^{xx}_{ij}=L^{xx}_{ji}$. Transport properties
can be defined in terms of $L_{ij}^{xx}$. For example, the electric conductivity $\sigma$,
thermoelectric power $S$, and the thermal conductivity $\kappa$
are
\begin{eqnarray}
\sigma &=& \frac{e^2}{T}L^{xx}_{11},\\
S &=& -\frac{1}{eT}\frac{L^{xx}_{12}}{L^{xx}_{11}},\\
\kappa &=& \frac{1}{T^2}\left(L^{xx}_{22}-\frac{(L^{xx}_{12})^2}{L^{xx}_{11}}\right).
\end{eqnarray}
In following context, we use $k_B=e=\hbar=1$. The practical value of $S$ is recovered
by multiplying the factor $k_B/e=86.3\mu V/K$, which we use as unit for thermopower.

In conventional thermoelectric problems, $L_{xx}^{ij}$ is theoretically defined
and experimentally measured at the DC limit. The extension to dynamical(frequency)
case is absent in standard textbooks but has been studied in detail
in Ref.~\cite{0034-4885-72-1-016501}. Here we give the outlines of the formalism.
Borrowed from Luttinger's derivation\cite{PhysRev.135.A1505}, an auxiliary
``gravitational'' field coupled to energy density is defined. An ``equivalence''
between the fictitious gravitational field and the temperature gradient is proved.
Then the transport coefficients $L^{xx}_{ij}$ can be written in terms of
correlation functions between particle current and(or) energy current. In
Ref.\cite{0034-4885-72-1-016501}, this formalism is generalized to temporally
and spatially periodic external fields, thus the transport coefficients become
momentum- and frequency-dependent functions, $L^{xx}_{ij}(\mathbf{q},\omega)$.

Some interesting remarks can be made on $L^{xx}_{ij}(\mathbf{q},\omega)$. The
thermodynamic limit corresponds to $\mathbf{q}\rightarrow 0$, and the static
fields correspond to the $\omega\rightarrow 0$ limit. The $L^{xx}_{ij}$ in
Eq.~(\ref{eq:linear_J1}) and Eq.~(\ref{eq:linear_J2}) can be approached by the
fast limit, i.e., taking $\mathbf{q}\rightarrow 0$ first and then $\omega\rightarrow 0$.
If we define the ``phase velocity'' of the external field, $v=\frac{\omega}{|q|}$,
the fast limit means $v\rightarrow\infty$, which gives the name ``fast''.
The slow limit means reserving this order, $\omega\rightarrow 0$ first and then
$\mathbf{q}\rightarrow$, thus $v\rightarrow 0$. The slow limit gives the Kelvin formula of thermopower
discussed in Ref.~\cite{PhysRevB.82.195105}. The high-frequency(AC) limit means
$\omega\rightarrow\infty$. In this case, we take the thermodynamic limit,
$\mathbf{q}\rightarrow 0$ first, and then $\omega\rightarrow \infty$. But from
the general formalism in Ref.~\cite{0034-4885-72-1-016501}, it can be shown
that the order of taking limits does not matter.

The dynamical transport coefficients with $\mathbf{q}\rightarrow 0$ are given by,
\begin{equation}
\label{eq:L_def}
L^{xx}_{ij}(\omega)=T\int_0^{\infty}dt e^{i(\omega+i0^+)t}\int_0^{\beta}d\tau\langle
J^{x}_j(-t-i\tau)J^{x}_i\rangle .
\end{equation}

For a given Hamiltonian $H$, the current operators are defined by following the
conservation laws\cite{mahan_many},
\begin{equation}
J^{x}_i=\frac{\partial O^{x}_i}{\partial t}=i[H, O^{x}_i].
\end{equation}
$O_i^x$ is the $x$-component of particle and heat polarization operator. Specifically,
\begin{eqnarray}
O_1^x &=& \sum_{i}R_i^x n_{i},\\
O_2^x &=& \sum_{i}R_i^x \left(h_{i}-\mu n_{i}\right),
\end{eqnarray}
where $n_{i}$ and $h_{i}$ are local particle and energy density operators.
The explicit forms of $n_i$ and $h_i$ are
determined by the Hamiltonian of specific models. In next subsection, we will
write $O_i$ and give $J_i$ for a general multiband model.

At DC limit, the imaginary part of $L_{ij}^{xx}(\omega=0)$ is zero, thus $S_0$
is determined by the real parts. For convenience, define
\begin{equation}
L_{ij}^0\equiv\re L_{ij}^{xx}(0),
\end{equation}
then we have
\begin{equation}
\label{eq:s_dc}
S_0\equiv\re S(\omega=0)=-\frac{1}{T}\frac{L_{12}^0}{L_{11}^0}.
\end{equation}

At AC limit, $L_{ij}^{xx}(\omega)$ is dominated by the imaginary part, with
a $O(1/\omega)$ leading order,
\begin{equation}
\label{eq:L_kk_relation}
\im L_{ij}^{xx}(\omega)=\frac{T}{\omega}L_{ij}^*+O(\frac{1}{\omega^2}).
\end{equation}
Using Lehnman's representation, it has been shown that $L_{ij}^*$
defined above is, up to a factor of $i$, the expectation values of commutators between
current and polarization operators\cite{PhysRevB.65.075102, 0034-4885-72-1-016501, PhysRevB.73.085117},
i.e.,
\begin{equation}
\label{eq:L_star_commu}
L_{ij}^*=i\langle [J_j^x, O_j^x] \rangle.
\end{equation}
Consequently, TEP at AC limit is
\begin{equation}
\label{eq:s_ac} S^*\equiv\re
S(\omega\rightarrow\infty)=-\frac{1}{T}\frac{L_{12}^*}{L_{11}^*}.
\end{equation}

$L_{ij}^*$ can be related to $\re L_{ij}(\omega)$. Applying Kramers-Kronig relation
and keeping the leading order in $1/\omega$, we have
\begin{equation}
\label{eq:L_star_sum}
L_{ij}^*=\frac{1}{\pi T}\int_{-\infty}^{\infty}d\omega \re L_{ij}^{xx}(\omega).
\end{equation}
Thus $L_{ij}^*$ is also connected to the sum rules of dynamical quantities.
For example, $L_{11}^*$ is proportional to the sum rule of conductivity\cite{pines, PhysRevLett.75.105}.
\begin{equation}
\label{eq:int_sigma}
L^*_{11}=\frac{2}{\pi}\int_0^{\infty}d\omega\re \sigma(\omega).
\end{equation}
Other sum rules are also derived in Ref.~\cite{0034-4885-72-1-016501} and \cite{PhysRevB.73.085117}.

\subsection{General formula of $L_{ij}^*$}\label{subsec:sum_rules}

Now we explicitly evaluate the commutator in Eq.~(\ref{eq:L_star_commu}) for a
general tight-binding Hamiltonian with local interaction, which will determine
the AC limit of TEP in this system. We start with the following Hamiltonian
\begin{eqnarray}
\label{eq:hamiltonian}
H &=& -\sum_{ij,\mu\nu}t_{ij}^{\mu\nu}c^{\dag}_{i\mu}c_{j\nu}+\sum_{i\mu}\epsilon_{\mu}c^{\dag}_{i\mu}c_{i\mu}\nonumber\\
{}&&+\sum_{i}\sum_{\alpha\beta\mu\nu}U_{\alpha\beta\mu\nu}c^{\dag}_{i\mu}c^{\dag}_{i\beta}c_{i\nu}c_{i\mu}.
\end{eqnarray}
$i$, $j$ are site indices. $\alpha$, $\beta$, $\mu$ and $\nu$ denote local orbitals.
$t_{ij}^{\mu\nu}$ is the hopping integral, and $U_{\alpha\beta\mu\nu}$ is the matrix element for
Coulomb interaction between local orbitals. $\epsilon_{\mu}$ is energy level of local
orbitals. The particle polarization operator is
\begin{equation}
O_1^x=\sum_{i}R_i^x\sum_{\mu}c^{\dag}_{i\mu}c_{i\mu},
\end{equation}
and the heat polarization operator is
\begin{eqnarray}
O_2^x &=& \sum_{i}R_{i}^x\left[-\frac{1}{2}\sum_{j,\mu\nu}\left(t_{ij}^{\mu\nu}c^{\dag}_{i\mu}c_{j\nu}
+t_{ji}^{\nu\mu}c^{\dag}_{j\nu}c_{i\mu}\right)\right.\nonumber\\
{}&&\left. + \sum_{\alpha\beta\mu\nu}U_{\alpha\beta\mu\nu}c^{\dag}_{i\mu}c^{\dag}_{i\beta}c_{i\nu}c_{i\mu}
+\sum_{\alpha}\left(\epsilon_{\alpha}-\mu\right)c^{\dag}_{i\alpha}c_{i\alpha}\right].\nonumber\\
\end{eqnarray}

The current operators turn out to be
\begin{eqnarray}
J_1^x &=& i[H,O_1^x]\nonumber\\
{} &=& -i\sum_{ij,\mu\nu}\left(R_j^x-R_i^x\right)t_{ij}^{\mu\nu}c^{\dag}_{i\mu}c_{j\nu},
\end{eqnarray}
and
\begin{eqnarray}
J_2^x&=&i[H, O_2^x]\nonumber\\
{}&=&\sum_{ijl, \mu\nu\alpha}\frac{i}{2}t_{il}^{\mu\alpha}t_{lj}^{\alpha\nu}\left(R_j^x-R_i^x\right)c^{\dag}_{i\mu}c_{j\nu}\nonumber\\
{}&&-\frac{i}{2}\sum_{ij,\alpha\beta}t_{ij}^{\alpha\beta}\left(R_j^x-R_i^x\right)\left(\epsilon_{\alpha}+\epsilon_{\beta}-2\mu\right)c^{\dag}_{i\alpha}c_{j\beta}\nonumber\\
{}&&-\frac{i}{2}\sum_{ij,\mu\nu}\left(R_j^x-R_i^x\right)\nonumber\\
{}&&\quad\times\left(\sum_{\alpha'\mu'\nu'}\left(U_{\nu\alpha'\mu'\nu'}-U_{\alpha'\nu\mu'\nu'}\right)c^{\dag}_{i\mu}c^{\dag}_{j\alpha'}c_{j\nu'}c_{j\mu'}\right.\nonumber\\
{}&&\qquad \left.+\sum_{\alpha'\beta'\nu'}\left(U_{\alpha'\beta'\mu\nu'}-U_{\alpha'\beta'\nu'\mu}\right)c^{\dag}_{i\alpha'}
c^{\dag}_{i\beta'}c_{i\nu'}c_{j\nu}\right).\nonumber\\
\end{eqnarray}

In the literature\cite{PhysRevB.67.115131}, $J_2^x$ is also written in a more compact form using the equation of motion in Heisenberg picture,
\begin{equation*}
J_2^x=-\frac{1}{2}\sum_{ij,\mu\nu}\left(R_j^x-R_i^x\right)t_{ij}^{\mu\nu}\left(\dot{c}^{\dag}_{i\mu}c_{j\nu}-c^{\dag}_{i\mu}\dot{c}_{j\nu}\right),
\end{equation*}
in which the dot means the time derivative,
\begin{equation*}
\dot{c}^{\dag}_{i\mu}=i[H, c^{\dag}_{i\mu}].
\end{equation*}

To compute $L_{11}^{*}$ and $L_{12}^*$, we need to further evaluate the commutators between
current operators and polarization operators. For $L_{11}^*$, this is simple and straightforward,
\begin{equation}
\label{eq:L_star_11}
L_{11}^*=\sum_{ij,\mu\nu}\left(R_j^x-R_i^x\right)^2\langle c^{\dag}_{i\mu}c_{j\nu} \rangle.
\end{equation}
However, $L_{12}^*$ leads to a complicate formula,
\begin{eqnarray}
\label{eq:L_star_12}
L_{12}^*{}&=&-\sum_{ijl, \mu\nu\alpha}\frac{1}{2}t_{il}^{\mu\alpha}t_{lj}^{\alpha\nu}\left(R_j^x-R_i^x\right)\langle c^{\dag}_{i\mu}c_{j\nu} \rangle\nonumber\\
{}&&+\frac{1}{2}\sum_{ij,\mu\nu}t_{ij}^{\mu\nu}\left(R_j^x-R_i^x\right)\left(\epsilon_{\mu}+\epsilon_{\nu}-2\mu\right)\langle c^{\dag}_{i\mu}c_{j\nu}\rangle\nonumber\\
{}&&+\frac{1}{2}\sum_{ij,\mu\nu}\left(R_j^x-R_i^x\right)\nonumber\\
{}&&\quad\times\left(\sum_{\alpha'\mu'\nu'}\left(U_{\nu\alpha'\mu'\nu'}-U_{\alpha'\nu\mu'\nu'}\right)\langle c^{\dag}_{i\mu}c^{\dag}_{j\alpha'}c_{j\nu'}c_{j\mu'}\rangle\right.\nonumber\\
{}&&\qquad \left.+\sum_{\alpha'\beta'\nu'}\left(U_{\alpha'\beta'\mu\nu'}-U_{\alpha'\beta'\nu'\mu}\right)\langle c^{\dag}_{i\alpha'}
c^{\dag}_{i\beta'}c_{i\nu'}c_{j\nu}\rangle \right).\nonumber\\
\end{eqnarray}
But this formula can be significantly simplified if we look at the equation of motion for the following
Greens's function,
\begin{equation}
G_{ji}^{\nu\mu}(\tau)=-\langle T_{\tau}c_{j\nu}(\tau)c^{\dag}_{i\mu} \rangle.
\end{equation}
$T_{\tau}$ is the time-ordering operator in imaginary time. Its equation of motion reads,
\begin{eqnarray*}
\frac{\partial G_{ji}^{\nu\mu}( \tau)}{\partial \tau} &=& \sum_{j'\nu'}t_{jj'}^{\nu\nu'}G_{j'j}^{\nu'\nu}(\tau)-\left(\epsilon_{\nu}-\mu\right)G_{ji}^{\nu\mu}(\tau)\nonumber\\
&&-\sum_{\alpha'\mu'\nu'}\left(U_{\alpha'\nu\mu'\nu'}-U_{\nu\alpha'\mu'\nu'}\right)\nonumber\\
&&\quad \times\langle T_{\tau}c^{\dag}_{j\alpha'}(\tau)c_{j\nu'}(\tau)c_{j\mu'}(\tau)c^{\dag}_{i\mu} \rangle.\nonumber\\
\end{eqnarray*}
Taking the $\tau\rightarrow 0^-$ limit leads to
\begin{multline}
\label{eq:Ucc}
\sum_{\alpha'\mu'\nu'}\left(U_{\nu\alpha'\mu'\nu'}-U_{\alpha'\nu\mu'\nu'}\right)\langle c^{\dag}_{i\mu}c^{\dag}_{j\alpha'}c_{j\nu'}c_{j\mu'}\rangle\\
=-\lim_{\tau\rightarrow 0^-}\frac{\partial G_{ji}^{\nu\mu}(\tau)}{\partial \tau}+\sum_{j'\nu'}t_{jj'}^{\nu\nu'}\langle c^{\dag}_{i\mu}c_{j'\nu'} \rangle\\
-\left(\epsilon_{\nu}-\mu\right)\langle c^{\dag}_{i\mu}c_{j\nu} \rangle.
\end{multline}
Substituting the last term in Eq.~(\ref{eq:L_star_12}) by the
right hand side of Eq.~(\ref{eq:Ucc}), we get
\begin{multline}
L_{12}^*=-\frac{1}{2}\sum_{ijl,\mu\nu\alpha}t_{il}^{\mu\alpha}t_{lj}^{\alpha\nu}\left[\left(R_j^x-R_i^x\right)^2\right.\\
\left.-\left(R_l^x-R_i^x\right)^2-\left(R_j^x-R_l^x\right)^2\right]\langle c^{\dag}_{i\mu}c_{j\nu} \rangle\\
-\sum_{ij,\mu\nu}t_{ij}^{\mu\nu}\left(R_j^x-R_i^x\right)^2\lim_{\tau\rightarrow 0^-}\frac{\partial}{\partial \tau}G_{ji}^{\nu\mu}(\tau).
\end{multline}

Using the fact that
\begin{equation*}
\langle c^{\dag}_{i\mu}c_{j\nu} \rangle=\lim_{\tau\rightarrow  0^-}G_{ji}^{\nu\mu}(\tau),
\end{equation*}
and performing Fourier transformation in both real space and imaginary time, we get
\begin{equation}
\label{eq:L_star_11_matsu}
L_{11}^*=\frac{1}{\beta}\sum_{\omega_n}e^{-i\omega_n 0^-}\sum_{k,\mu\nu}\left(\frac{\partial^2\epsilon_k^{\mu\nu}}{\partial k_x^2}\right)G_{k}^{\nu\mu}(i\omega_n),
\end{equation}
and,
\begin{multline}
\label{eq:L_star_12_matsu}
L_{12}^*=\frac{1}{\beta}\sum_{\omega_n}e^{-i\omega_n 0^-}\sum_{k,\mu\nu}\left[\sum_{\alpha}\left(\frac{\partial \epsilon_{k}^{\mu\alpha}}{\partial k_x}\right)
\left(\frac{\partial \epsilon_{k}^{\alpha\nu}}{\partial k_x}\right)\right.
\left.+i\omega_n\left(\frac{\partial^2\epsilon_k^{\mu\nu}}{\partial k_x^2}\right)\right]G_k^{\nu\mu}(i\omega_n).
\end{multline}
$\epsilon_k^{\mu\nu}$ is Fourier transformation of hopping amplitudes,
\begin{equation}
\epsilon_k^{\mu\nu}=-\sum_{R}e^{ikR}t^{\mu\nu}(R),
\end{equation}
where we have utilized the translational invariance,
\begin{equation}
t_{ij}^{\mu\nu}=t^{\mu\nu}(R_j-R_i).
\end{equation}

It is straightforward to convert the Matsubara summation to the integration in real frequencies.
\begin{equation}
\label{eq:L_star_11_real}
L_{11}^*=\int_{-\infty}^{\infty} d\omega\sum_{k,\mu\nu}\left(\frac{\partial^2\epsilon_k^{\mu\nu}}{\partial k_x^2}\right)f(\omega)A_k^{\nu\mu}(\omega),
\end{equation}
and
\begin{multline}
\label{eq:L_star_12_real}
L_{12}^*=\int_{-\infty}^{\infty} d\omega\sum_{k,\mu\nu}\left[\sum_{\alpha}\left(\frac{\partial \epsilon_{k}^{\mu\alpha}}{\partial k_x}\right)
\left(\frac{\partial \epsilon_{k}^{\alpha\nu}}{\partial k_x}\right)\right.
\left.+\omega\left(\frac{\partial^2\epsilon_k^{\mu\nu}}{\partial k_x^2}\right)\right]f(\omega)A_k^{\nu\mu}(\omega).
\end{multline}
$f(\omega)=1/(1+\exp(\beta\omega))$ is the Fermi function. $A_k^{\nu\mu}(\omega)=-\frac{1}{\pi}G_k^{\nu\mu}(\omega)$ is
the spectral function.

Eq.~(\ref{eq:L_star_11_matsu}), Eq.~(\ref{eq:L_star_12_matsu}), Eq.~(\ref{eq:L_star_11_real})
and Eq.~(\ref{eq:L_star_12_real}) are main results in this work. They are derived from a
general formalism of dynamical thermoelectric transport outline in Sec.~\ref{subsec:gen_form}
and a multiband Hamiltnian, Eq.~(\ref{eq:hamiltonian}). The equation of motion is exact
and no approximation is assumed in the derivation. These equations indicate that $L_{11}^*$
and $L_{12}^*$, and thus $S^*$ are determined by the non-interacting band structure and the
single-particle spectral fundtion.

\section{$S_0$ and $S^*$ in a one-band Hubbard model}\label{sec:dmft}

In this section, we discuss $S_0$ and $S^*$ of one-band Hubbard model
in the scenario of dynamical mean field theory(DMFT), using the formalism we presented in
previous sections.

The Hamiltonian of one-band Hubbard model is
\begin{equation}
\label{eq:one_band_hubbard}
H=-\sum_{ij, \sigma}t_{ij}c^{\dag}_{i\sigma}c_{j\sigma}+U\sum_{i}n_{i\uparrow}n_{i\downarrow}.
\end{equation}
In DMFT, it is mapped to a single-impurity Anderson
model\cite{PhysRevB.45.6479} supplemented by the self-consistent
condition, which reads,
\begin{equation}
\label{eq:self_cons}
\frac{1}{i\omega_n+\mu-\Delta(i\omega_n)-\Sigma(i\omega_n)}=\sum_k
G_k(i\omega_n).
\end{equation}
On the left hand side is the local Green's function on the impurity. $\Delta(i\omega_n)$
is the hybridization function of the impurity model. On the
right hand side, $G_k(i\omega_n)$ is the Green's function of lattice electrons,
\begin{equation*}
G_k(i\omega_n)=\frac{1}{i\omega_n+\mu-\epsilon_k-\Sigma(i\omega_n)},
\end{equation*}
with $\epsilon_k$ the non-interacting dispersion relation of the lattice model,
and $\Sigma(i\omega_n)$ the self energy for both local and lattice
Green's function in the self-consistent condition. In DMFT, both coherent
and incoherent excitations in a correlated metal are treated on the
same footing\cite{RevModPhys.68.13}.

In DMFT, the evaluation of transport coefficients,e.g., Eq.~(\ref{eq:L_def}), can
be significantly simplified. Because the $k$-dependence falls solely on the
non-interacting dispersion $\epsilon_k$, the vertex corrections vanishes\cite{PhysRevB.47.3553}.
Consequently, $\re L_{ij}(\omega)$ can be written in terms of
single-particle spectral function in real frequency.
\begin{eqnarray}
\label{eq:Lw_real} \re L_{ij}(\omega) &=& \pi
T\sum_{k,\sigma}\left(\frac{\partial \epsilon_k}{\partial
k_x}\right)^2\int_{\infty}^{\infty}
d\omega'(\omega'+\frac{\omega}{2})^{i+j-2}\nonumber\\
{}&&\times\left(\frac{f(\omega')-f(\omega'+\omega)}{\omega}\right)
A_k(\omega')A_k(\omega'+\omega).\nonumber\\
\end{eqnarray}
Notice that here the dependence of $\re L_{ij}(\omega)$ on the single-particle
spectral function is generally approximate for
a finite-dimensional system, which is achieved due to the vanishing of vertex
corrections exact only in infinite dimensions. But the dependence of $L_{ij}^*$
on single-particle spectral function is exact, as pointed out at the end of
Sec.~\ref{subsec:gen_form}.

Another question is on the sum rule of the approximate $\re L_{ij}(\omega)$,
i.e., if we substitute Eq.~(\ref{eq:Lw_real}) into the definition of $L_{ij}^*$,
Eq.~(\ref{eq:L_star_sum}), wether or not it will give the same form of $L_{ij}^*$
as we have derived in last section. The answer to this question is yes and we
a brief proof for this one-band case in the Appendix but the extension to
multiband case is straightforward. This means that ignoring vertex correction
will modify the distribution of weight in $\re L_{ij}(\omega)$, but will not
change the integrated weight.

The DC limit of $\re L_{ij}(\omega)$, $L^0_{ij}$ can be obtained by takeing the
limit $\omega\rightarrow 0$, which gives,
\begin{equation}
\label{eq:L_re_0} L^0_{ij}=\pi
T\sum_{k,\sigma}\left(\frac{\partial
\epsilon_k}{\partial k_x}\right)^2 \int_{-\infty}^{\infty}d\omega
\omega^{i+j-2}\left(-\frac{\partial f(\omega)}{\partial
\omega}\right)A_k(\omega)^2.
\end{equation}

Therefore in the framework of DMFT, $S_0$ is computed from Eq.~(\ref{eq:L_re_0}).
The AC limit, $S^*$ can be computed from Eq.~(\ref{eq:L_star_11_matsu}) and Eq.~(\ref{eq:L_star_12_matsu}),
or Eq.~(\ref{eq:L_star_11_real}) and Eq.~(\ref{eq:L_star_11_real}). In principle,
Matsubara frequency and integration over real frequency give identical results.
But in practice, especially in numerical computations on correlated systems,
correlation functions in Matsubara frequencies are more easily accessible. For example,
among various impurity solvers in DMFT, quantum Monte Carlo method(QMC), i.e., Hirsch-Fye
method\cite{PhysRevLett.56.2521} and recently developed continuous time QMC\cite{PhysRevLett.97.076405,
PhysRevB.75.155113} are implemented in imaginary time. To get correlation functions in
real frequencies, numerical realization of analytical continuation has to be
employed, such as maximum entropy method, which is a involved procedure and usually
special care has to be taken of. In this case, a formulae in Matsubara frequencies
will significantly simplify the calculation.

Due to the bad convergence of the series,
Eq.~(\ref{eq:L_star_11_matsu}) and Eq.~(\ref{eq:L_star_12_matsu}) are not
appropriate for direct implementation into numerical computations.
Following standard recipe(separating and analytically evaluating the badly
convergent part), we transform them in a form more friendly to numerics. For
the one-band Hubbard model,
\begin{equation}
\label{eq:L_star_11_num}
L^*_{11}=\sum_{k,\sigma}\left(\frac{\partial^2\epsilon_{k\sigma}}{\partial
k_x^2}\right)\left(\frac{1}{\beta}\sum_{\omega_n}\re
G_k(i\omega_n)-\frac{1}{2}\right),
\end{equation}
and
\begin{multline}
\label{eq:L_star_12_num}
L_{12}^*=\sum_{k,\sigma}\left(\frac{\partial\epsilon_k}{\partial
k_x}\right)^2\frac{1}{\beta}\sum_{\omega_n}\re G_k(i\omega_n)
\times\left[1+2\omega_n \im G_k(i\omega_n)\right].
\end{multline}

\subsection{Low temperature limit. }\label{subsec:lowT}

At low temperatures(low-T), the derivative of Fermi function,
$(-\partial f(\omega)/\partial\omega)$ in the integrand of
Eq.~(\ref{eq:L_re_0}) becomes Dirac-$\delta$ function-like, thus only
the low energy part of the spectral weight near Fermi surface contributes to the integral.
The low energy part of the self energy of a Fermi liquid $\Sigma(\omega)$
can be approximated by a Taylor expansion in terms of $\omega$ and
$T$.
\begin{eqnarray*}
\re \Sigma(\omega) &\simeq& \left(1-\frac{1}{Z}\right)\omega,\\
\im \Sigma(\omega) &\simeq& \frac{\gamma_0}{Z^2}(\omega^2+\pi^2 T^2)+\frac{1}{Z^3}(a_1\omega^2+\omega T^2).
\end{eqnarray*}
Previous studies\cite{PhysRevLett.80.4775, haule} showed that at low-T limit, $L^0_{11}\propto
Z^2/T$ and $L^0_{12}\propto ZT$, thus
$S_0=-L^0_{12}/(TL^0_{11})\propto T/Z$.

Since we are interested in the relation between $S_0$ ad $S^*$, it
would be convenient to write $L^*_{12}$ and $L^*_{11}$ in terms of
the conventional transport function, $(\partial\epsilon_k/\partial
k_x)^2$. This can be achieved by performing integration by part on
the summation over $k$ in in Eq.~(\ref{eq:L_star_11_real})
and Eq. ~\ref{eq:L_star_12_real}, then we have
\begin{equation*}
\label{eq:L_star_decomp} L^*_{ij}=L^*_{ij,I}+L^*_{ij,II},
\end{equation*}
with
\begin{eqnarray}
L^*_{ij,I} &=& \sum_{k,\sigma}\left(\frac{\partial
\epsilon_k}{\partial k_x}\right)^2\int d\omega \left(-\frac{\partial
f(\omega)}{\partial\omega}\right)\nonumber\\
{}&&
\times\omega^{i+j-2}\left(-\frac{1}{\pi}\right)\im\left[G_k(\omega)
Z(\omega)\right],\label{eq:L_star_I}\\
L^*_{ij,II} &=& \sum_{k,\sigma}\left(\frac{\partial
\epsilon_k}{\partial k_x}\right)^2\int d\omega
f(\omega)\left(-\frac{1}{\pi}\right)\nonumber\\
{} &&
\times\im\left[G(\epsilon,\omega)\frac{\partial}{\partial\omega}
\left(\omega^{i+j-2}(1-Z(\omega))\right)\right],\nonumber\\
\label{eq:L_star_II}
\end{eqnarray}
where we have defined
\begin{equation*}
Z(\omega)=\frac{1}{1-\partial\Sigma(\omega)/\partial\omega}.
\end{equation*}

We introduced the function $Z(\omega)$, which is dependent on the
derivative of self energy with respect to energy $\omega$. The
integrand in $L^*_{ij,I}$(Eq.(\ref{eq:L_star_I})) also has
the derivative of Fermi function. Also
notice that at low-T, $Z(\omega=0)=Z$, which is the renormalization
factor of correlated Fermi liquid. Then $L^*_{ij,I}$ resembles $L^0_{12}$
except for the power of $\im G_k(\omega)$. Low temperature expansion
show that $L^*_{11,I}\propto Z$, and $L^*_{12,I}\propto T^2$.
Therefore, if $L^*_{11,II}$ and $L^*_{12,II}$ were absent,
$S^*=-(TL^*_{12,I})/L^*_{11,I}\propto T/Z$, which is similar to the
low-T behavior of $S_0$.

However, $L^*_{11,II}$ and $L^*_{12,II}$ do not vanish in general at
low-T limit. First, at low-T limit, the integral over $\omega$ in
Eq.~(\ref{eq:L_star_II})
\begin{equation*}
\int_{-\infty}^{\infty}d\omega f(\omega)\quad \text{is replaced
by}\quad \int_{-\infty}^{0}d\omega.
\end{equation*}
Then both the real and imaginary part of $G_k(\omega)$ and
$Z(\omega)$ below Fermi surface have to contribute to the leading
order of $L^*_{ij,II}$, unless $\Sigma(\omega)$ is independent, or
at least weakly dependent on $\omega$, leading $Z(\omega)\simeq 1$,
and then the integrand in $L^*_{ij,II}$ would vanish. But this in
general can not be true. For example, in a correlated Fermi liquid
phase near the Mott transition of Hubbard model, $\Sigma(\omega)$
contains the information of coherent quasiparticles at Fermi surface
as well as that of incoherent excitations in high-energy Hubbard
bands, thus $\Sigma(\omega)$ will depend on $\omega$ in very
different ways at these separated energy scales. At low energy
scale, $Z(\omega\simeq0)\simeq Z$, and $Z$ is significantly less
than $1$ near Mott transition.
Therefore, at low-T limit, $L^*_{ij,II}$ will exhibit a finite value
at low-T limit. So the total value of $L^*_{12}$ will be dominated
by $L^*_{12,II}$ instead of the $\sim T^2$ contribution from
$L^*_{12,I}$. The finiteness of $L^*_{11}$ can be also justified by
the general sum rule Eq.~(\ref{eq:L_star_11_real}), which indicates
that $L^*_{11}$ is proportional to the kinetic energy.
Consequently, $S^*$ will diverge $1/T$-like at low-T limit for a
correlated Fermi liquid.

There are some circumstances in which $Z(\omega)=1$ and
$L^*_{ij,II}$ vanishes. One example is that in a static mean field
theory, such as Hartree-Fock approximation, $\Sigma(\omega)$ is
independent on $\omega$, thus in static mean field theory, it is
possible that $S^*$ can show a similar behavior to that of $S_0$
at low temperature.

\subsection{High temperature limit.}\label{subsec:highT}

In the literature, the high temperature limit of thermopower\cite{PhysRevB.13.647},
or known as Mott-Heikes formulor, has been widely used
as a benchmark for thermoelectric capability\cite{PhysRevB.62.6869}
for correlated materials. Here we discuss the high temperature limit of $S^*$
implied from the formulae we have derived.

The high temperature limit relevant for correlated systems was approached
by first taking the limit $U\rightarrow\infty$, which excludes the double-
occupancy in hole-doped systems or the vacancy in electron-doped systems,
then taking the high temperature limit $T\rightarrow 0$. This leads to
two major simplification. First, by definition in thermodynamics,
\begin{equation*}
\frac{\mu}{T}=-\left(\frac{\partial s}{\partial N}\right)_{E,V}.
\end{equation*}
Here $s$ is the entropy and $N$ is number of electrons. $s$ can be calculated
by counting all possible occupation states satisfying the $U\rightarrow\infty$
limit. It turns out that $\frac{\mu}{T}$ is a constant determined by the electron
density. Thus $\mu$ is proportional to $T$ at high temperature.
The second simplification is that at high temperature, we can approximate
the single particle spectral function by a rigid band picture, namely,
\begin{equation}
\label{eq:rigid_band}
\tilde{A}_k(\omega)=A_k(\omega-\mu).
\end{equation}
$\tilde{A}_k(\omega)$ is a function of $\omega$ but independent of temperature
and chemical potential. Applying these simplification to Eq.~(\ref{eq:L_star_11_real})
and Eq.~(\ref{eq:L_star_12_real}), and keeping the leading order in $T$, we have
\begin{eqnarray*}
L_{11}^* &=& \frac{1}{1+e^{-\beta\mu}}\int d\omega\sum_{k,\sigma}\left(\frac{\partial^2\epsilon_k}{\partial k_x^2}\right)\tilde{A}_k(\omega),\\
L_{12}^* &=& \frac{-\mu}{1+e^{-\beta\mu}}\int d\omega\sum_{k,\sigma}\left(\frac{\partial^2\epsilon_k}{\partial k_x^2}\right)\tilde{A}_k(\omega).
\end{eqnarray*}
Therefore, at high temperature limit,
\begin{equation}
S^*=-\frac{L_{12}^*}{TL_{11}^*}=\frac{\mu}{T}.
\end{equation}
This is the same result to the high temperature limit of $S_0$ in Ref.~\cite{PhysRevB.13.647}.
Thus the leading order of $S^*$ is identical to the leading order of $S_0$
at high temperature.

\section{Numerical results}\label{sec:numerical}

In this section, we compute the dynamical thermoelectric power
$S(\omega)$ by dynamical mean field theory(DMFT).
We use exact diagonalization(ED) as the impurity solver.
The advantage of the ED solver is the Green's functions
can be computed simultaneously in real and Matsubara
frequencies. Thus we have two approaches to compute the AC limit $S^*$.
The first one is to substitute the Green's function in
Matsubara frequencies into Eq.~(\ref{eq:L_star_11_matsu}) and
Eq.~(\ref{eq:L_star_12_matsu}). The second method starts from computing
$\re L_{11}(\omega)$ and $\re L_{12}(\omega)$ from spectral functions
$A_k(\omega)$ using Eq.~(\ref{eq:Lw_real}) for a wide range of $\omega$,
Kramers-Kronig relation is implemented to compute $\im L_{11}(\omega)$
and $\im L_{12}(\omega)$, and finally with the value of $L_{11}^*$ and
$L_{12}^*$ obtained by fitting Eq.~(\ref{eq:L_kk_relation}) at the
$\omega\rightarrow\infty$ limit. The second method is more laborious
but here we use it as a check for our formulae in Matsubara frequencies.

We study one-band Hubbard model on square and triangular
lattices and consider only the hopping between nearest neighboring sites.


\subsection{Square lattice}\label{subsec:square}

In this section, we compute the thermoelectric transport
coefficients and thermoelectric power for a hole-doped Hubbard
model on square lattice. We use the bandwidth $D$ as the unit for
frequency $\omega$, temperature $T$ and interaction strength $U$.
For square lattice, $D=8|t|$, $t$ is the hopping constant.

\begin{figure}

\begin{tabular}{cc}
\includegraphics[angle=270,width=\figwidth]{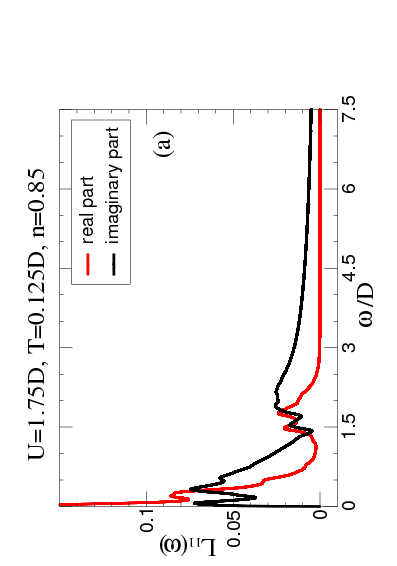}& \includegraphics[angle=270,width=\figwidth]{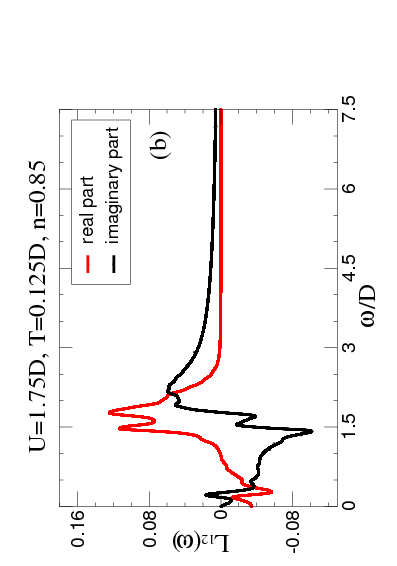}\\
\includegraphics[angle=270,width=\figwidth]{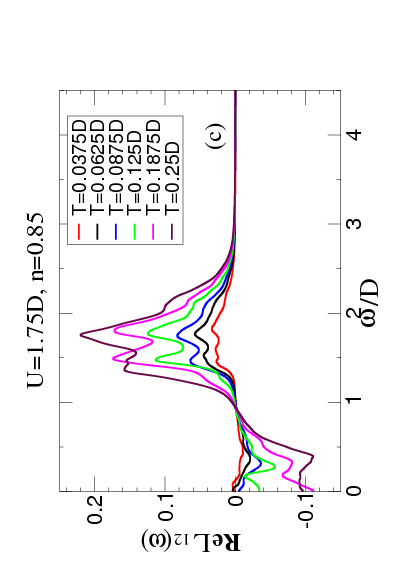}& \includegraphics[angle=270,width=\figwidth]{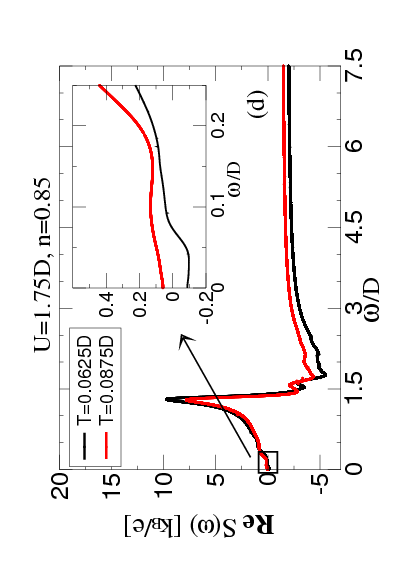}\\
\end{tabular}
\caption{\label{fig:freq_depend} Frequency-dependent transport
coefficients and thermoelectric power of a hole-doped one-band Hubbard model
on square lattice. $U=1.75D$ and $n=0.85$. (a) $\re L_{11}(\omega)$
and $\im L_{11}(\omega)$ at $T=0.125D$. (b) $\re L_{12}(\omega)$ and
$\im L_{12}(\omega)$ at $T=0.125D$. (c) The evolution of $\re
L_{12}(\omega)$ with temperature. (d) $\re S(\omega)$ at $T=0.0625D$
and $T=0.0875D$. The inset blows up the region near $\omega=0$.}
\end{figure}

In Fig.~\ref{fig:freq_depend} we show the frequency-dependent
quantities for $U=1.75D$ and $n=0.85$. Fig.~\ref{fig:freq_depend}-(a)
and -(b) show the thermoelectric transport coefficients
$L_{11}(\omega)$ and $L_{12}(\omega)$ by their real(red line) and
imaginary part(black line). The real parts are computed from
Eq.~(\ref{eq:Lw_real}). The imaginary parts are computed
from Kramers-Kronig relation. Three contributions are recognizable in
$\re L_{11}$: i) The low frequency peak due to transition within the
resonance peak of quasiparticles. ii) The transition between
quasiparticles and the lower Hubbard band, which accounts for the
hump at $\omega\sim 0.5D$. iii) The weight around $\omega\sim U$,
which is due to the incoherent excitations between Hubbard bands.
Same features also exist in $\re L_{12}(\omega)$, but the feature
near $\omega\sim 0$\, i.e., transition between quasiparticles,
and the transition between quasiparticles and lower Hubbard band,
are much less obvious. This is because the DC limit $L^0_{12}$
is dominated by the particle-hole asymmetry of the band velocity
$\partial\epsilon_k/\partial k_x$ and the spectral function
$A_k(\omega)$, due to the $\omega^{i+j-2}=\omega$ term in the
integrand of Eq.~(\ref{eq:L_re_0}) for $L^0_{12}$. Thus at small $\omega$,
$\re L_{12}(\omega)$ is significantly impaired, compared to $\re L_{11}(\omega)$. Therefore the
transition by incoherent excitations around $\omega\sim U$ takes a major part in the total
weight in $\re L_{12}(\omega)$, and the sum
rule of $\re L_{12}(\omega)$, i.e., $L^*_{12}$, is also dominated by the
incoherent excitations. $\im L_{11}(\omega)$ and $\im
L_{12}(\omega)$ are odd functions of $\omega$ and vanish at
$\omega=0$. It is evident that the real parts
approach to zero much faster than the imaginary parts at AC limit($\omega\rightarrow\infty$),.
Fig.~\ref{fig:freq_depend}-(c) shows the evolution of $\re
L_{12}(\omega)$ as temperatures. The dominance of the incoherent
excitations is robust as the variation of temperature.
Fig.~\ref{fig:freq_depend}-(d) shows the real part of thermoelectric
power, $\re S(\omega)$ for $T=0.0625D$ and $T=0.0875D$. The inset blows
up the region near $\omega=0$, indicating that $S_0$ displays $+$
or $-$ signs at different temperatures.

\begin{figure}

\begin{tabular}{cc}
\includegraphics[angle=270,width=\figwidth]{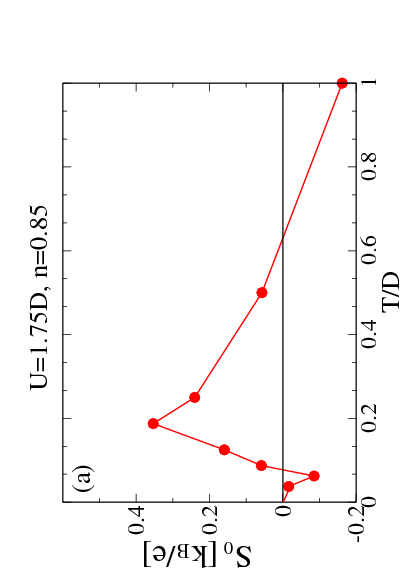} &
\includegraphics[angle=270,width=\figwidth]{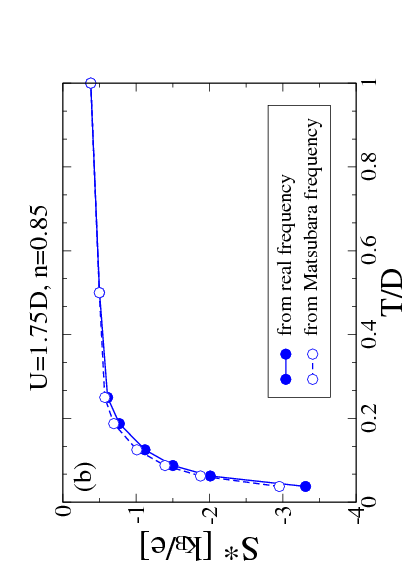} \\
\end{tabular}
\caption{\label{fig:temp_dep}  (a): Temperature dependence of
$S_0$. (b): Temperature dependence of $S^*$ obtained from real frequencies (filled circles) and from Matsubara
frequencies (open circles). Square lattice.}
\end{figure}

In Fig.~\ref{fig:temp_dep}-(a) and (b) we show $S_0$ and $S^*$ at
various temperatures. On the one side, in Fig.~\ref{fig:temp_dep}-(a), $S_0$ presents
multiple changes of sign with temperature increased. The sign change at
lower temperature($T\sim 0.1D$) demonstrates the crossover from the low-temperature hole-like
coherent quasiparticles to incoherent excitations at intermediate temperature.
Around $T=0.2D$, where $S_0$ reaches its maximum positive value, where the coherent
quasiparticles have almost diminished. The second sign change
around $T=0.6D$ indicates a subtle competition between the spectral weight of
lower and higher Hubbard band. As temperature increases, the asymmetry
between the two Hubbard bands near Fermi surface becomes less significant because more spectral
weight from the higher Hubbard band takes part into the transport and the sign
of $S_0$ is determined by the difference between the weight of lower and
higher Hubbard. This crossover is thus considered to be responsible for the
second sign change\cite{PhysRevB.65.075102} and also has been observed experimentally
\cite{PhysRevB.54.17469}. Therefore, above $T=0.6D$, the transport is completely
dominated by incoherent excitations from both Hubbard bands. On the other side, in Fig.~\ref{fig:temp_dep}-(b),
the situation for $S^*$ is quite different. $S^*$ does not change sign and keeps
negative in the shown temperature range. Towards low temperature, $S^*$ blows up,
consistent with our argument based on a Fermi liquid self energy in Sec.~\ref{subsec:lowT}.
Towards high temperature, i.e., when the temperature is well above the coherence
regime, $S_0$ and $S^*$ have the same sign and similar magnitude. We notice that $S_0$ in
Fig.~\ref{fig:temp_dep}-(a) does not converge to the value predicted by the
Mott-Heikes formula in the correlated regime($S_{MH}\simeq 1.04 k_B/e$, from Eq.~(11)
in Ref.~\cite{PhysRevB.13.647}). This is because in our case, with $U=1.5D$, the
requirement for $|t|\ll T\ll U$ can not be satisfied for a wide range of temperature.
Thus at high temperature, e.g., when $T>0.6D$, the states with double occupancy
can not be excluded and they are responsible for the second sign change
in $S_0$ as discussed above.

In Fig.~\ref{fig:temp_dep}-(b), we show $S^*$ obtained by the two methods mentioned at
the beginning of Sec.~\ref{sec:numerical}. The solid circles represents $S^*$ by
fitting $\im L_{11}(\omega)$ and $\im L_{12}(\omega)$ in real frequency at $\omega\rightarrow \infty$
limit. The open circles represent $S^*$ computed using Eq.~(\ref{eq:L_star_11_matsu})
and Eq.~(\ref{eq:L_star_12_matsu}). The values of $S^*$ at open and closed circles
are very close, indicating the consistency between the real and Matsubara frequency
approach to calculate $S^*$.

\begin{figure}
\includegraphics[angle=270,width=0.8\columnwidth]{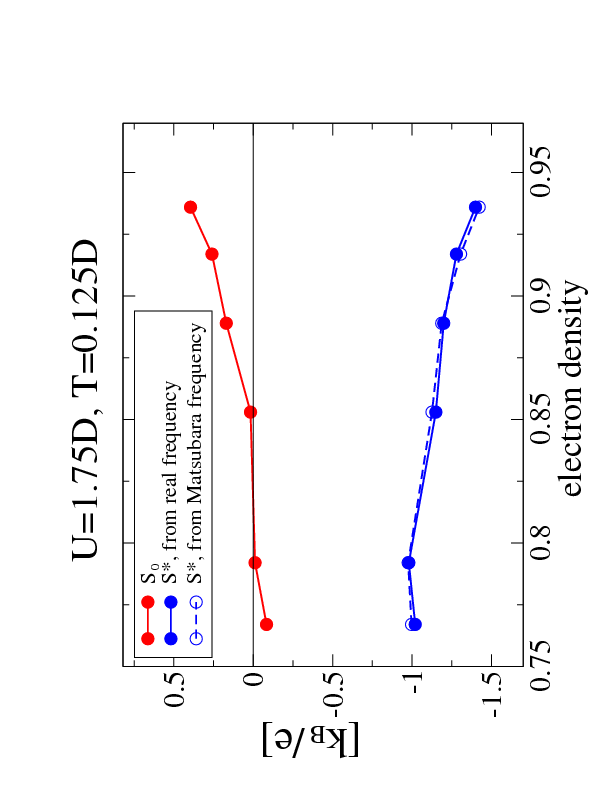}
\caption{\label{fig:dop_dep}  Doping dependence
of $S_0$ and $S^*$ for $U=1.75D$. $S^*$ was obtained from real frequencies
(filled circles) and Matsubara frequencies (open circles).
The temperature here is $T=0.125D$. Square lattice. }
\end{figure}

The dependence on electron density of $S_0$ and $S^*$ is more non-trivial, which is difficult to tell
from analytical formulas. Fig.~\ref{fig:dop_dep} shows $S_0$ and $S^*$ at various
densities for $U=1.75D$. $S_0$ changes sign from positive at half filling to negative
as electron density decreases, while $S^*$ remains negative. The behavior of $S_0$ here is
also due to the breakdown of coherence as the evolution of spectral weight.
In a doped Mott insulator, the quasiparticle peak gradually diminishes as
the system is doped away from half-filling\cite{PhysRevB.53.16214}. Thus near
half-filling, the transport is dominated by the coherent excitations near Fermi
surface. But when the doping is heavy enough to kill quasiparticles, transport
is carried by incoherent excitations in the Hubbard bands. Therefore $S_0$ turns
to a same sign with $S^*$, since $S^*$ is dominated by the Hubbard bands(see
Fig.~\ref{fig:freq_depend}-(c) and discussion there). In Fig.~\ref{fig:dop_dep}
we also put the results of $S^*$ by real and Matsubara frequency approach.

\subsection{Triangular lattice}\label{subsec:triangular}

Recent interest on thermoelectric performance of correlated systems
was attributed to the discovery of TEP enhancement
in highly electron doped cobaltates\cite{Nat.Mat.5.537.}.
The $Co$ atoms in the $Co O_2$ layers form a triangular lattice.
The physics behind the large TEP in $Na_x Co O_2$ is highly non-trivial. For
example, the $Na$ potential is crucial to induce the correlation in
$Na_{0.7}Co O_2$\cite{PhysRevLett.98.176405}, and the spin and orbital degrees
of freedom are argued to be a key factor for the enhancement
\cite{PhysRevB.62.6869, Nat.423.425}. These complexities are
beyond a single band Hubbard on a triangular lattice. Here we
only focus on some qualitative features of $S_0$ and $S^*$ in a
electron-doped single band Hubbard model on triangular lattice.

In triangular lattice, $U=12|t|$ and we use a positive t. $S^*$ in this section is solely
computed by using Eq.~(\ref{eq:L_star_11_matsu}) and Eq.~(\ref{eq:L_star_11_matsu}).

\begin{figure}
\includegraphics[angle=270,width=0.85\columnwidth]{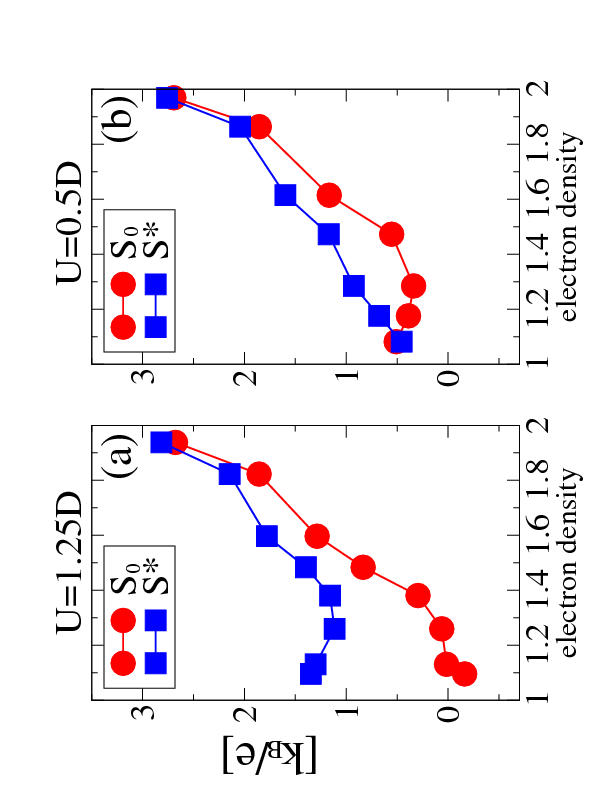}
\caption{\label{fig:s_triang} (a) and (b): Density dependence of
$S_0$ and $S^*$ for $U=1.25D$ and $U=0.5D$.
$S^*$ was obtained from the Matsubara frequencies. Triangular lattice.}
\end{figure}

Fig.~\ref{fig:s_triang}-(a) and -(b) shows the density dependence of
$S_0$ and $S^*$ for two different interaction strength. Here we present
the full range for electron doping. Here $S^*$ is from the summation over
Matsubara frequency. For $U=1.25D$(Fig.~\ref{fig:s_triang}-(a)), $S_0$ is negative near
half-filling and changes to positive after a small amount of doping.
As the density approaches to band insulator($n=2$), the merging of $S_0$ and $S^*$
is very evident. For smaller interaction strength, i.e., $U=0.5D$, $S_0$ and
$S^*$ also display similar trend through the range of electron density. This
behavior is similar to the case on square lattice, Fig.~\ref{fig:dop_dep}.
The discrepancy between $S_0$ and $S^*$ is most evident for $U=1.25D$ and near
half-filling($n=1.0$), since around this regime the coherent quasiparticles
take a significant role in transport. For electron density larger than $1.5$,
which is the range of interest for cobaltate, the trend of $S^*$ shows that
it is a reasonable approximation to $S_0$.

\section{Summary}\label{sec:conclusion}
Using the formulae derived in Sec.~\ref{sec:formal}, we investigate
to what extent the AC limit of thermoelectric power, $S^*$, can be
a reasonable approximation to the DC limit, $S_0$. Analytical and
numerical results on a single-band Hubbard model show that below and
around coherent temperature, i.e., when the spectral weight around
quasiparticle peak dominates in the thermoelectric transport, the
behaviors of $S_0$ and $S^*$ are significantly different. Specifically,
$S_0$ displays multiple sign changes around the coherent temperature,
but $S^*$ does not. But when the temperature is well beyond the coherent
regime, thus the transport properties are dominated by the incoherent
excitations, $S^*$ shows same sign and similar magnitude to $S_0$ and can
give reasonable prediction on the behavior of $S^*$.

Our work suggest that a realistic implementation of Eq.~(\ref{eq:L_star_11_matsu})
and Eq.~(\ref{eq:L_star_12_matsu}) in LDA+DMFT codes can serve as a useful guide
for the search of high performance thermoelectric materials among the strongly
correlated electron systems, which have a very broad temperature regime
characterized by incoherent transport.

At the time of writing, we are aware of a recent work by M. Uchida et al.\cite{uchida},
in which the incoherent thermoelectric transport over a wide temperature
range is studied in a typical density-driven Mott transition system $La_{1-x}Sr_xVO_3$
and the validity of Mott-Heikes formula for real strongly correlated materials
is verified.

\section{Acknowledgement}
This work was supported by the NSF under NSF grant DMR-0906943.
CW was supported by the Swiss Foundation for Science (SNF).
Useful discussions with K. Haule and V. Oudovenko are gratefully
acknowledged.

\appendix
\section{Sum rules for $\re L_{12}(\omega)$ and $\re L_{11}(\omega)$ in DMFT}
\label{app2:dmft_sum_rule}

In this appendix, we compute $L_{11}^*$ and $L_{12}^*$ in the framework
of dynamical mean field theory and show they also obey the general formulae,
Eq.~(\ref{eq:L_star_11_real} and Eq.~(\ref{eq:L_star_12_real}).

In terms of retarded current-current correlations,
\begin{equation}
\re L_{12}^{xx}(\omega)=-\frac{1}{\omega}\im \left[ \int_{-\infty}^{\infty}dt
e^{i(\omega+i0^+)}\left[-i\theta(t)\langle [J_j(t), J_i] \rangle\right]\right],
\end{equation}
which can be computed in Matsubara frequencies by standard diagrammatic
techniques\cite{mahan_many}. In the infinite dimension limit, a significant
simplification is achieved because all nonlocal irreducible vertex
collapse and only the first bubble diagram survives
\cite{PhysRevB.47.3553, PhysRevLett.75.105}.
This simplification leads to
\begin{eqnarray}
\label{eq:app2_Lw} \re L^{xx}_{ij}(\omega) &=& \pi
T\sum_{k,\sigma}\left(\frac{\partial \epsilon_k}{\partial
k_x}\right)^2\int_{\infty}^{\infty}
d\omega'\left(\omega'+\frac{\omega}{2}\right)^{i+j-2}\nonumber\\
{}&&\times\left(\frac{f(\omega')-f(\omega'+\omega)}{\omega}\right)
A_k(\omega')A_k(\omega'+\omega).\nonumber\\
\end{eqnarray}
Now we calculate $L^*_{ij}$. Using Eq.~(\ref{eq:L_star_sum}),
\begin{multline}
L^*_{12}=\sum_{k,\sigma}\left(\frac{\partial \epsilon_k}{\partial
k_x}\right)^2\int d\omega
d\omega'\left(\omega'+\frac{\omega}{2}\right)\\
\times\left(\frac{f(\omega')-f(\omega+\omega')}{\omega}\right)A_k(\omega')A_k(\omega'+\omega).
\end{multline}

Changing variables by
\begin{eqnarray*}
\omega_1&=&\omega+\omega',\nonumber\\
\omega_2&=&\omega,\nonumber
\end{eqnarray*}
leads to
\begin{multline}
\label{eq:app2_L_star_12}
L^*_{12}=\sum_{k,\sigma}\left(\frac{\partial\epsilon_{k}}{\partial
k_x}\right)^{2}\int
d\omega_{1}d\omega_{2}f(\omega_{2})A_{k}(\omega_{1})A_{k}(\omega_{2})\\
+2\sum_{k,\sigma}\left(\frac{\partial\epsilon_{k}}{\partial k_x}\right)^{2}\int
d\omega_{1}d\omega_{2}\frac{\omega_{2}}{\omega_{1}-\omega_{2}}f(\omega_{2})A_{k}(\omega_{1})A_{k}(\omega_{2}).
\end{multline}
The sum rule $\int d\omega_1 A_k(\omega_1)=1$ simplifies the first
term to
\begin{equation*}
\sum_{k,\sigma}\left(\frac{\partial\epsilon_{k}}{\partial k_x}\right)^{2}\int
d\omega_{2}f(\omega_{2})A_{k}(\omega_{2}).
\end{equation*}
In the second term, Kramer-Kronig relation
can be used to eliminate the integral over $\omega_1$, i.e.,
\begin{equation*}
\int d\omega_1\frac{A_k(\omega_1)}{\omega_1-\omega_2}=-\re
G_{k}(\omega_2).
\end{equation*}
Then we use the fact that
\begin{equation*}
2\re G_k(\omega) \im G_k(\omega)=\im G^2_k(\omega)
\end{equation*}
and
\begin{equation*}
\frac{\partial}{\partial
k_x}G_k(\omega)=G_k^2(\omega)\frac{\partial\epsilon_k}{\partial k_x},
\end{equation*}
to simplify the second term on the right hand side of
Eq.~(\ref{eq:app2_L_star_12}) to
\begin{equation*}
\sum_{k,\sigma}\left(\frac{\partial \epsilon_k}{\partial
k_x}\right)\int d\omega_2 \omega_2 f(\omega_2)
\left(\frac{1}{\pi}\right)\frac{\partial}{\partial k_x}\im
G_k(\omega_2).
\end{equation*}
Applying integration by part over $k$, it turns out to be
\begin{equation*}
\sum_{k,\sigma}\left(\frac{\partial^2\epsilon_k}{\partial
k_x^2}\right)\int d\omega_2 \omega_2 f(\omega_2)A_k(\omega_2).
\end{equation*}
Combined with the first term, we have
\begin{equation}
L^*_{12}=\sum_{k,\sigma}\int
d\omega\left(\left(\frac{\partial\epsilon_{k}}{\partial
k_x}\right)^{2}+\omega\left(\frac{\partial^2 \epsilon_k}{\partial
k_x^2}\right)\right)f(\omega)A_k(\omega).
\end{equation}

The calculation for $L^*_{12}$ is similar and straightforward, which
results in
\begin{equation}
L^*_{11}=\sum_{k,\sigma}\int d\omega\left(\frac{\partial^2
\epsilon_k}{\partial k_x^2}\right)f(\omega)A_k(\omega).
\end{equation}

\bibliography{ref_wcg}

\end{document}